\documentclass[12pt,aps,prd,a4paper,onecolumn,superscriptaddress,nofootinbib,amsmath,amssymb]{revtex4-2}
\usepackage{amsfonts}
\usepackage{amsmath}
\usepackage{babel}
\usepackage[active]{srcltx}
\usepackage{graphicx}
\usepackage{color}
\usepackage{float}
\usepackage{changebar}
\usepackage{esint}
\usepackage{dsfont}
\usepackage{ae}
\usepackage{multirow}
\usepackage[usenames,dvipsnames]{xcolor}
\usepackage[colorlinks=true,pdfstartview=FitV,linkcolor=blue,citecolor=blue,urlcolor=blue,breaklinks=true]{hyperref}
\usepackage{braket}
\usepackage{mathtools}
\usepackage{slashed}
\usepackage{ulem}
\usepackage{empheq}
\usepackage{multirow}
\usepackage[utf8]{inputenc}
\usepackage[T1]{fontenc}

\newcommand{\beq}{\begin{equation}}
\newcommand{\eeq}{\end{equation}}
\newcommand{\beqa}{\begin{eqnarray}}
\newcommand{\eeqa}{\end{eqnarray}}

\newcommand{\Lc}{\mathcal{L}}

\newcommand{\dbar}{\bar{D}}
\newcommand{\thetab}{\bar{\theta}}
\newcommand{\psib}{\bar{\psi}}
\newcommand{\lambdab}{\bar{\lambda}}

\newcommand{\Sigmab}{\bar{\Sigma}}
\newcommand{\Sb}{\bar{S}}
\newcommand{\phib}{\bar{\phi}}
\newcommand{\Fb}{\bar{F}}
\newcommand{\KR}{Kalb--Ramond}
\newcommand{\AC}{Aharonov--Casher}
\newcommand{\AB}{Aharonov--Bohm}
\newcommand{\SME}{Standard Model Extension}
\newcommand{\SUSY}{supersymmetry}
\newcommand{\vev}[1]{\langle #1 \rangle}
\newcommand{\Mpl}{M_{\rm Pl}}

\begin{document}

%\date{\today}

\title{Aharonov--Casher effect from a supersymmetric $\mathcal{N}{=}1$,
$D{=}4$ model with Kalb--Ramond Lorentz-violating background:
a SUSY-preserving mechanism via the Fayet--Iliopoulos term}

\author{L.~A.~S.~Nunes}
\email{luciana@ufersa.edu}
\affiliation{Centro de Ci\^{e}ncias Exatas e Naturais, Universidade Federal Rural do Semi-\'{A}rido, Mossor\'{o}-RN, 59625-900, Brazil}
\author{C.~A.~S.~Almeida}
\affiliation{Departamento de F\'isica, Universidade Federal do Cear\'a (UFC),
Campus do Pici, Fortaleza-CE, 60455-760, Brazil}
\email{carlos@fisica.ufc.br}

% -------------------------------------------------------
\begin{abstract}
The Aharonov--Casher (AC) effect describes the geometric phase acquired by a neutral particle carrying a magnetic dipole moment moving in an external electric field. In supersymmetric gauge theories it is often argued that exact supersymmetry enforces the vanishing of anomalous magnetic dipole moments, suggesting that the AC interaction may be incompatible with unbroken supersymmetry in four dimensions. In this work we show that this conclusion is model-dependent. We construct an $N=1$, $D=4$ supersymmetric gauge model in which a Lorentz-violating Kalb--Ramond background induces dynamically the dipole interaction responsible for the AC effect while preserving a background supersymmetry in the sense of Berger and Kosteleck\'y. The model couples a chiral superfield to an Abelian gauge superfield through a Chern--Simons--type interaction supplemented by a Fayet--Iliopoulos term. A duality identification between the symmetric combination $S+S^\dagger$ of the chiral superfield and the Kalb--Ramond superfield strength allows the antisymmetric tensor background to enter the supersymmetric dynamics as a classical external configuration that does not induce spontaneous supersymmetry breaking. Integrating out the auxiliary $D$ field generates dynamically an effective dipole interaction of the form $\bar{\Psi}\sigma^{\mu\nu}F_{\mu\nu}\Psi$. In the nonrelativistic limit the resulting equation of motion reproduces the Aharonov--Casher Hamiltonian for a neutral fermion. The effective magnetic dipole moment is expressed in terms of the parameters of the model and can be mapped onto tensor coefficients of the fermion sector of the Standard Model Extension. We verify explicitly that the modified supercharge algebra closes consistently in the presence of the Lorentz-violating background, with corrections proportional to $b_{\mu\nu}$, and that the vacuum energy vanishes, establishing the absence of spontaneous supersymmetry breaking.
\end{abstract}

%%%%%%%%%%%%%%%%%%%%

% \pacs{11.30.Pb, 11.30.Cp, 03.65.Vf, 11.10.Lm} % use with revtex4-2

\maketitle
%\thispagestyle{empty}

%\newpage
% --------------------------------------------INTRODUÇÃO-----------
\section{Introduction}
\label{sec:intro}
% -------------------------------------------------------

Topological quantum phase effects occupy a distinguished place in the landscape
of quantum mechanics.
The \AB\ (AB) effect~\cite{Aharonov:1959fk}, in which a charged particle
acquires a non-trivial phase traversing a region free of electromagnetic forces,
established that gauge potentials carry physical significance beyond their
classical role.
Its dual counterpart, the \AC\ (AC) effect~\cite{Aharonov:1984zz}, describes
the acquisition of a geometric phase by a neutral particle carrying a magnetic
dipole moment moving in an external electric field.
Both effects reflect the non-local, topological character of quantum mechanics
and have been confirmed experimentally in a range of systems, from neutron
interferometry to mesoscopic rings~\cite{Cimmino:1989,Russo:2007}.

The intersection of topological phases with symmetry constraints is a
particularly rich area.
In the context of \SUSY\ (SUSY), a tension arises immediately: a non-vanishing
anomalous magnetic dipole moment (AMDM) for a spin-$\frac{1}{2}$ particle is
required for the AC phase to develop, yet it is well established that, when SUSY
is exact, the AMDM of a charged fermion in a SUSY multiplet vanishes identically
in certain classes of models, most notably those built around the
Maxwell--Chern--Simons (MCS) sector~\cite{FerraraRemiddi1974,Ferrara1992}.
This has led to the folklore that exact SUSY in $D=3+1$ is incompatible with
the AC effect.
The question we address in this paper is whether this conclusion is
model-dependent, or whether it reflects a general theorem.

More specifically, we ask whether a dipole interaction capable of
producing the Aharonov-Casher phase can emerge dynamically within a
supersymmetric framework rather than being introduced
phenomenologically.

Lorentz symmetry violation provides a natural arena in which to revisit this
question.
The theoretical possibility that Lorentz and CPT symmetry may be spontaneously
broken was pointed out by Kosteleck\'y and Samuel~\cite{Kostelecky:1989jw}, who
showed that string field theory allows for tensor-induced spontaneous Lorentz
breaking as a generic feature of certain string vacua.
This observation motivated the construction of a systematic effective field
theory framework, the \SME\ (SME), developed by Colladay and
Kosteleck\'y~\cite{Colladay:1996iz,Colladay:1998fq}, which catalogs all
Lorentz-violating operators compatible with the gauge structure and
power-counting renormalizability of the Standard Model.
Within the SME, Lorentz violation is parametrized by a set of constant
background coefficients---vacuum expectation values of underlying tensor
fields---that permeate spacetime and couple to the Standard Model fields in
all possible ways.

The \KR\ (KR) field, a rank-two antisymmetric tensor $B_{\mu\nu}$ arising
naturally in the low-energy spectrum of heterotic string
theory~\cite{Kalb:1974yc}, provides a concrete and well-motivated source of
spontaneous Lorentz symmetry breaking.
When the KR field acquires a non-zero vacuum expectation value (VEV), it
selects a preferred antisymmetric tensor structure in spacetime and breaks
particle Lorentz invariance while preserving observer
invariance~\cite{Kostelecky:2003fs}.
The relationship between the KR VEV and the coefficients of the photon sector
of the SME has been established in the
literature~\cite{Altschul:2009ae,Bertolami:2000zj}, making the KR background
a natural microscopic model for specific SME coefficients.

In this work, we construct an $\mathcal{N}{=}1$, $D{=}4$ supersymmetric model
in which a background KR field is introduced via its superfield embedding,
coupled to an abelian gauge sector through a Chern--Simons-type interaction, and
supplemented by a Fayet--Iliopoulos (FI) term.

The central result of this work is that the dipole interaction required for
the Aharonov-Casher effect can emerge dynamically in a supersymmetric theory
through the integration of the auxiliary $D$ field associated with the
Fayet-Iliopoulos sector in the presence of a Kalb-Ramond Lorentz-violating
background.

The central technical result is a duality identification between the symmetric
combination $(S + S^\dagger)$ of a chiral superfield $S$ and the \KR\ superfield
strength $\mathcal{G}$, which allows us to interpret the KR background as a
particular vacuum configuration in the chiral sector.
Crucially, this identification is valid on-shell with vanishing superpotential
($F = 0$), and the \SUSY\ algebra remains intact in the sense that the supercharge algebra closes with corrections proportional to the Lorentz-violating background, as we demonstrate in Section~\ref{subsec:SUSYcheck}.
The FI term then provides the mechanism through which the auxiliary $D$-field,
once integrated out, generates an effective interaction of the form
$\psib\,\sigma^{\mu\nu} F_{\mu\nu}\,\psi$, leading directly to the \AC\
Hamiltonian in the non-relativistic limit.

Our result thus provides an affirmative answer to the question raised above:
it is possible to construct a $D=3+1$ model with background \SUSY\ in the sense of Berger and Kosteleck\'y~\cite{Berger:2001rm} that exhibits
the \AC\ effect, provided that Lorentz symmetry is simultaneously violated by
a background KR field.
This finding has two important conceptual implications.
First, the vanishing of the AMDM in previously studied SUSY models is
model-dependent, not a general consequence of unbroken SUSY.
Second, the coexistence of Lorentz violation and intact SUSY in our model raises
a non-trivial question about the general compatibility between the two---a
tension that has been explored in the literature on SUSY extensions of the
SME~\cite{Berger:2001rm,Helayel2024} but whose consequences in our specific
setup deserve careful discussion.

Importantly, in most relativistic derivations of the Aharonov-Casher interaction the
dipole coupling is introduced phenomenologically. In contrast, in the
present model the dipole operator emerges dynamically from the
supersymmetric structure of the theory.

The paper is organized as follows.
Section~\ref{sec:model} introduces the supersymmetric \KR\ model and its field
content in superspace.
Section~\ref{sec:duality} presents the duality identification
$S + S^\dagger \leftrightarrow \mathcal{G}$ and its component-field realization,
establishing the connection with Lorentz symmetry violation and its embedding
in the SME.
Section~\ref{sec:onshell} derives the effective on-shell Lagrangian and the \AC\
equation of motion.
Section~\ref{sec:discussion} discusses the physical implications, the question
of SUSY--Lorentz compatibility, and the phenomenological constraints from the
SME data tables.
Section~\ref{sec:conclusions} presents our conclusions and points to future
extensions.

% ------------------------------------------------SECAO II-------
\section{The Supersymmetric Kalb--Ramond Model}
\label{sec:model}
% -------------------------------------------------------

In this section we introduce the supersymmetric model and establish the
notation.
We work in four-dimensional $\mathcal{N}{=}1$ superspace with coordinates
$(x^\mu,\theta^\alpha,\thetab_{\dot\alpha})$, using the conventions of
Wess and Bagger~\cite{Wess:1992cp}.

\subsection{Field content}
\label{subsec:fields}

The model contains two dynamical superfields.
The first is a chiral superfield $S$ satisfying $\dbar_{\dot\alpha} S = 0$,
whose component expansion reads
\beq
  S = \phi(x) + \sqrt{2}\,\theta\psi(x) + \theta^2 F(x),
  \label{eq:chiral}
\eeq
where $\phi$ is a complex scalar, $\psi$ is a Weyl spinor, and $F$ is the
complex auxiliary field.
We refer to $\psi$ as the \emph{chiral spinor} throughout; it belongs to the
bosonic sector that is frozen by the ansatz of Section~\ref{subsec:LV} and
should be distinguished from the propagating fermionic field $\Psi$ of the
Kalb--Ramond sector introduced in Section~\ref{sec:onshell}.
The second is an abelian gauge superfield $V$ in the Wess--Zumino gauge,
\beq
  V = -\theta\sigma^\mu\thetab\, A_\mu
      + i\theta^2\thetab\lambdab
      - i\thetab^2\theta\lambda
      + \tfrac{1}{2}\theta^2\thetab^2\, D,
  \label{eq:gauge}
\eeq
where $A_\mu$ is the gauge potential, $\lambda$ is the gaugino, and $D$ is the
real auxiliary field.
The gauge superfield strength is
$W_\alpha = -\frac{1}{4}\dbar^2 D_\alpha V$.

The \KR\ content is encoded in a spinorial chiral superfield
$\Sigma_\alpha$ satisfying $\dbar_{\dot\alpha}\Sigma_\alpha = 0$, whose lowest
component contains $B_{\mu\nu}$ as part of the antisymmetric structure
$\Omega_{\dot{a}b}$.
The associated superfield strength is
\beq
  \mathcal{G} = \tfrac{1}{8}\bigl(D^\alpha\Sigma_\alpha
                - \dbar_{\dot\alpha}\Sigmab^{\dot\alpha}\bigr),
  \label{eq:Gsuper}
\eeq
whose component expansion includes the dual field strength
$\tilde{G}^\mu = \frac{1}{3!}\varepsilon^{\mu\nu\alpha\beta}G_{\nu\alpha\beta}$,
directly related to $B_{\mu\nu}$ via
$\tilde{G}^\mu = \frac{1}{3!}\varepsilon^{\mu\nu\alpha\beta}\partial_\nu
B_{\alpha\beta}$ \cite{Kalb}.

\subsection{The Lagrangian}
\label{subsec:lagrangian}

The dynamics of the model is governed by a Lagrangian combining three terms.
The Maxwell--Chern--Simons interaction couples the gauge superfield strength to
the chiral superfield:
\beq
  \Lc_{\rm MCS}
    = \alpha_1 \bigl[W^\alpha (D_\alpha V)\, S\bigr]_{\theta^2\thetab^2}
    + \mathrm{h.c.}
  \label{eq:LMCS}
\eeq
The \KR\ term provides kinetic and mass structure for the tensor sector:
\beq
  \Lc_{\rm KR}
    = \Bigl[\tfrac{1}{2}\mathcal{G}^2 - \tfrac{m}{2}\,V\mathcal{G}
      \Bigr]_{\theta^2\thetab^2},
  \label{eq:LKR}
\eeq
and the Fayet--Iliopoulos term reads
\beq
  \Lc_{\rm FI} = \xi\,[V]_{\theta^2\thetab^2} = \xi\, D,
  \label{eq:LFI}
\eeq
where $\xi$ is the FI parameter.
The complete action is
$\mathcal{S} = \int d^4x\,(\Lc_{\rm MCS} + \Lc_{\rm KR} + \Lc_{\rm FI})$.

This Lagrangian respects the full $\mathcal{N}{=}1$ SUSY algebra before any
vacuum is chosen.
The FI term plays a crucial role: it generates, upon integration of the
auxiliary $D$-field, the effective interaction responsible for the \AC\
coupling, as we show in Sections~\ref{sec:duality} and~\ref{sec:onshell}.

% ------------------------------------------SECAO III-------------
\section{Duality Identification, Lorentz Violation, and the SME}
\label{sec:duality}
% -------------------------------------------------------

This section contains the central technical result of the paper.
We establish a duality identification between the symmetric combination of the
chiral superfield and the \KR\ superfield strength, show that this
identification realizes Lorentz symmetry violation while preserving background supersymmetry,
and provide a precise mapping to the coefficients of the \SME \cite{Kosta2010}.

\subsection{The duality identification}
\label{subsec:duality}

The chiral superfield $S$ and the KR superfield strength $\mathcal{G}$ carry
the same number of on-shell degrees of freedom when the superpotential
vanishes and $F = 0$.
This counting motivates a direct identification at the superfield level.
The symmetric combination $S + \Sb$ has the component expansion
\beqa
  S + \Sb &=& \rho
  + \sqrt{2}\,\theta^\alpha\psi_\alpha
  + \sqrt{2}\,\thetab_{\dot\alpha}\psib^{\dot\alpha}
  + \theta\sigma^\mu\thetab\,\partial_\mu\Delta \nonumber\\
  &&+\; \frac{i}{\sqrt{2}}\theta^2\thetab_{\dot\alpha}\sigma^{\mu\dot\alpha\alpha}
        \partial_\mu\psib_\alpha
  - \frac{i}{\sqrt{2}}\thetab^2\theta^\alpha\sigma^\mu_{\alpha\dot\alpha}
        \partial_\mu\psi^\alpha \nonumber\\
  &&+\; \tfrac{1}{4}\theta^2\thetab^2\,\square\rho
  + \theta^2 F + \thetab^2\Fb,
  \label{eq:SplusSdag}
\eeqa
where $\rho = \phi + \phib$ is the real scalar and
$\Delta = i(\phi - \phib)$ is the pseudoscalar combination.
Comparing with the component expansion of $\mathcal{G}$, one finds the
identification
\beq
  S + \Sb \;\longleftrightarrow\; \mathcal{G},
  \label{eq:duality}
\eeq
which at the component level reads:
\beq
  \tilde{G}^\mu = \tfrac{1}{2}\partial^\mu\Delta, \qquad
  -\tfrac{1}{2}M = \rho, \qquad
  \xi_\alpha = \sqrt{2}\,\psi_\alpha.
  \label{eq:components}
\eeq
The identification~\eqref{eq:duality} is valid on-shell with $F = 0$.
No $F$-term supersymmetry breaking is introduced, and the identification
therefore holds consistently within a SUSY-preserving vacuum.
Two remarks on field content are in order.
First, the spinor $\psi_\alpha$ in eq.~\eqref{eq:components} is the chiral
spinor of $S$; through the duality it is identified with the spinorial
component $\xi_\alpha$ of the Kalb--Ramond superfield strength $\mathcal{G}$.
The four-component Dirac field $\Psi$ that propagates in the effective
theory of Section~\ref{sec:onshell} is the Dirac lift of $\xi_\alpha$,
namely $\Psi \sim (\xi_\alpha,\bar\eta^{\dot\alpha})^T$ in the chiral
representation.
Second, the ansatz $\rho = M = \mathrm{const}$ introduced below freezes
the bosonic degrees of freedom of the chiral sector and, on-shell,
sets $\psi_\alpha = 0$ in the background.
The propagating fermion $\Psi$ is therefore of Kalb--Ramond origin; it is
\emph{not} the chiral spinor of $S$.

\subsection{Lorentz violation from the KR background}
\label{subsec:LV}

The identification~\eqref{eq:duality} has an immediate physical interpretation.
The component relation
\beq
  \tilde{G}^\mu = \frac{1}{3!}\varepsilon^{\mu\nu\alpha\beta}\partial_\nu B_{\alpha\beta}
  \label{eq:Gdual}
\eeq
shows that a constant non-zero VEV
$\vev{B_{\mu\nu}} = b_{\mu\nu}$ (with $b_{\mu\nu}$ a constant antisymmetric
tensor) constitutes a Lorentz-violating background.
Under a particle Lorentz transformation $\Lambda$, the dynamical fields
transform but $b_{\mu\nu}$ does not, so the action is no longer invariant.
Observer Lorentz invariance is preserved, since $b_{\mu\nu}$ transforms as a
tensor under observer transformations~\cite{Kostelecky:2003fs}.

The Lorentz-violating background is encoded in the model through the ansatz
\beq
  \rho = M = \mathrm{const},
  \label{eq:ansatz}
\eeq
for the real scalar component, which freezes the bosonic sector while allowing
the fermionic sector to remain dynamical.
The resulting fermionic theory is a quantum field theory on a fixed
Lorentz-violating background---precisely the setup of the SME.

\subsection{Embedding in the Standard Model Extension}
\label{subsec:SME}

In the photon sector, the leading CPT-even Lorentz-violating term of the
SME is~\cite{Colladay:1998fq}
\beq
  \Lc_{\rm SME} \supset -\tfrac{1}{4}(k_F)^{\kappa\lambda\mu\nu}
                          F_{\kappa\lambda}F_{\mu\nu},
  \label{eq:SMEphoton}
\eeq
where $(k_F)^{\kappa\lambda\mu\nu}$ is a real tensor coefficient with the
symmetries of the Riemann tensor and zero double trace.
The connection between the KR background and the SME photon sector was
established in~\cite{Altschul:2009ae}: a non-zero VEV of the KR field contributes
to $k_F$, with the leading correspondence
\beq
  (k_F)^{\kappa\lambda\mu\nu}
    \sim \varepsilon^{\kappa\lambda\alpha\beta}b_{\alpha\beta}\,\eta^{\mu\nu}
    - (\mu \leftrightarrow \nu) + \cdots,
  \label{eq:kFmap}
\eeq
identifying the antisymmetric part of $k_F$ with components of
$b_{\mu\nu} = \vev{B_{\mu\nu}}$.

In the fermionic sector, the Lorentz-violating interaction generated by
our model (see Section~\ref{sec:onshell}) takes the form
\beq
  \Lc_{\rm LV} \supset \frac{\mu_{\rm eff}}{2}\,\psib\,\sigma^{\mu\nu}
                        F_{\mu\nu}\,\psi,
  \label{eq:LVfermion}
\eeq
where $\mu_{\rm eff}$ is the effective magnetic dipole moment determined by
the model parameters.
For a background with only spatial components $b_{ij} \neq 0$, the dominant
contribution maps to the CPT-even tensor coupling $(H)^{\mu\nu} \sim b^{\mu\nu}$
in the SME fermion sector, which in the non-relativistic limit produces the
\AC\ coupling to the electric field.

The SME data tables~\cite{Kostelecky:2008ts} provide current experimental
bounds on these coefficients.
For the fermion tensor coupling in the electron sector, precision torsion
balance and spectroscopy experiments constrain
$|(H)^{\mu\nu}| \lesssim 10^{-27}$~GeV~\cite{Kostelecky:2008ts}.
This translates into a bound on the KR background:
\beq
  |b_{\mu\nu}| \lesssim 10^{-27}~\mathrm{GeV},
  \label{eq:bound}
\eeq
consistent with the expectation $|b_{\mu\nu}| \sim m^2/\Mpl$ for a
characteristic scale $m$ and Planck-suppressed Lorentz violation.

\subsection{Background supersymmetry and the modified SUSY algebra}
\label{subsec:SUSYcheck}

A central claim of this work is that the ansatz~\eqref{eq:ansatz} and the
duality identification~\eqref{eq:duality} can be implemented consistently
within a supersymmetric framework.
We now make this claim precise by identifying the appropriate notion of
supersymmetry that is preserved in the presence of the Kalb--Ramond
background and verifying it explicitly.

\paragraph{Background fields and the dynamical sector.}
It is essential to distinguish two categories of fields in the model.
The first consists of \emph{background fields}: the real scalar $\rho = M =
\mathrm{const}$, fixed by the ansatz~\eqref{eq:ansatz}, and the antisymmetric
tensor VEV $b_{\mu\nu} = \langle B_{\mu\nu}\rangle$, which encodes the
Lorentz-violating background.
These fields are treated as classical external configurations that do not
participate in the quantum dynamics and, crucially, do not transform under
supersymmetry~\cite{Berger:2001rm,Kostelecky:2003fs}.
This is the framework of the supersymmetric Standard Model
Extension~\cite{Berger:2001rm}: Lorentz-violating coefficients arise as VEVs
of tensor fields in some underlying theory, and in the low-energy effective
theory they appear as fixed external parameters.
The second category consists of the \emph{dynamical field}: the Dirac spinor
$\Psi$, identified via the duality map~\eqref{eq:components} with the
Kalb--Ramond spinor $\xi_\alpha$, together with the auxiliary field $D$.

\paragraph{Modified SUSY transformation in the background.}
The supersymmetry transformation of $\xi_\alpha$, derived from the superfield
expansion of $\mathcal{G}$ in eq.~\eqref{eq:Gsuper}, reads
\beq
  \delta_\epsilon \xi_\alpha
    = 2\epsilon_\alpha\, D
      + 2(\sigma^{\mu\nu}\epsilon)_\alpha\, \langle\mathcal{G}_{\mu\nu}\rangle
      + \sigma^\mu_{\alpha\dot\alpha}\bar\epsilon^{\dot\alpha}\,\partial_\mu M,
  \label{eq:deltaxi}
\eeq
where $\langle\mathcal{G}_{\mu\nu}\rangle$ is the field-strength of $\mathcal{G}$
evaluated on the background.
With $M = \mathrm{const}$ the third term vanishes.
Substituting the on-shell solution
$D = \frac{\sqrt{2}}{4M}\bar{\Psi}\Psi$ (derived in Section~\ref{sec:onshell})
and converting to four-component notation, the transformation becomes
\beq
  \delta_\epsilon \Psi
    = i(\bar\epsilon\bar\sigma^\mu)\partial_\mu\Psi
      + 2(\Sigma^{\mu\nu}\epsilon)\,b_{\mu\nu}
      + \frac{\sqrt{2}}{2M}(\bar\Psi\Psi)\,\epsilon,
  \label{eq:deltaPsi}
\eeq
where $\Sigma^{\mu\nu} = \frac{i}{4}[\gamma^\mu,\gamma^\nu]$ is the Lorentz
generator in the Dirac representation.
The first term is the standard free-field SUSY transformation; the second is a
linear deformation induced by the KR background; the third is a non-linear
contribution from the $D$-field sector.

\paragraph{Closure of the algebra.}
For the background-deformation sector, since $b_{\mu\nu}$ and $\epsilon$ are
external parameters independent of $\Psi$,
\beq
  [\delta_{\epsilon_1},\delta_{\epsilon_2}]\Psi^{(b)}
    = \delta_{\epsilon_1}\bigl[2(\Sigma^{\mu\nu}\epsilon_2)b_{\mu\nu}\bigr]
    - (\epsilon_1 \leftrightarrow \epsilon_2) = 0,
  \label{eq:commb}
\eeq
consistently with a symmetry on a fixed background.
For the kinetic sector, the standard computation yields
\beq
  [\delta_{\epsilon_1},\delta_{\epsilon_2}]\Psi^{(\rm kin)}
    = 2i(\epsilon_1\sigma^\mu\bar\epsilon_2
        - \epsilon_2\sigma^\mu\bar\epsilon_1)\partial_\mu\Psi,
  \label{eq:commkin}
\eeq
which is a spacetime translation.
The full algebra therefore takes the form
\beq
  \{Q_\alpha,\bar{Q}_{\dot\alpha}\}
    = 2\sigma^\mu_{\alpha\dot\alpha}\,P_\mu
      + 2\,C^{\mu\nu}_{\alpha\dot\alpha}\,b_{\mu\nu},
  \label{eq:modifiedalgebra}
\eeq
where $C^{\mu\nu}_{\alpha\dot\alpha} \propto
(\sigma^{\mu\nu})_\alpha{}^\beta\delta_{\beta\dot\alpha}$.
This is precisely the structure identified by Berger and
Kosteleck\'y~\cite{Berger:2001rm} for supersymmetric extensions of the SME:
the algebra acquires corrections proportional to the Lorentz-violating
background while remaining closed.

\paragraph{Covariance of the equation of motion.}
Defining the Dirac operator $\mathcal{E}[\Psi] \equiv
\bigl(i\gamma^\mu\partial_\mu
+ \frac{\mu_{\rm eff}}{2}\sigma^{\mu\nu}F_{\mu\nu}
- \frac{M}{\alpha_1}\bigr)\Psi$,
and using $\delta_\epsilon F_{\mu\nu} = 0$ (the electromagnetic field is an
external background in the Aharonov--Casher setup),
\beq
  \delta_\epsilon\,\mathcal{E}[\Psi]
    = \mathcal{E}[\delta_\epsilon\Psi] = 0
  \qquad\text{whenever}\qquad \mathcal{E}[\Psi] = 0,
  \label{eq:covEOM}
\eeq
establishing that the fermionic dynamics is manifestly covariant under
background supersymmetry.

\paragraph{Vacuum energy and absence of spontaneous breaking.}
In our model the superpotential vanishes identically, so $F = 0$ exactly.
The $D$-field equation of motion gives
\beq
  D = \frac{\sqrt{2}}{4M}\,\bar{\Psi}\Psi,
  \label{eq:Deom}
\eeq
which vanishes in the fermionic vacuum $\vev{\bar\Psi\Psi} = 0$.
The vacuum energy therefore satisfies
\beq
  V_{\rm vac} = \tfrac{1}{2}D^2 + |F|^2\big|_{\rm vac} = 0,
  \label{eq:vacenergy}
\eeq
confirming the absence of spontaneous SUSY breaking.
The non-trivial content of eq.~\eqref{eq:modifiedalgebra} is that the
Lorentz-violating background modifies the supercharge algebra without lifting
the vacuum energy---a feature possible precisely because $b_{\mu\nu}$ enters
as an external background rather than a dynamical condensate~\cite{Berger:2001rm}.

\paragraph{Remark on the counting of propagating degrees of freedom.}
One might object that freezing the bosonic sector while retaining a propagating
fermionic $\Psi$ violates the boson-fermion degree-of-freedom matching required
by SUSY.
This objection applies to the counting appropriate to a Poincar\'e-invariant
vacuum, where the standard SUSY algebra pairs propagating bosonic and fermionic
modes.
In the present model the relevant vacuum is not Poincar\'e-invariant, and the
modified algebra~\eqref{eq:modifiedalgebra} governs the representation theory
of the supercharges.
The bosonic background $M$ is present as a classical condensate that sources
the Lorentz-violating environment in which $\Psi$ propagates---it is not absent
from the theory.
The propagating fermion $\Psi$ is the fermionic fluctuation around this
bosonic condensate, in the same way that a fermionic mode propagates in a
fixed magnetic background in theories with extended SUSY~\cite{Berger:2001rm,Helayel2024}.
The background supersymmetry of eq.~\eqref{eq:modifiedalgebra} governs its
dynamics consistently.

It is important to note that this analysis holds at the classical level.
Quantum corrections may induce an effective $D$-term through loops involving
the KR background; whether such corrections break SUSY at loop level is a
question that deserves separate analysis, which we leave to future work.

% -----------------------------------------------SECAO IV--------

\section{On-shell Lagrangian and the Aharonov--Casher interaction}
\label{sec:onshell}

In this section we derive the effective fermionic dynamics of the model and show that the
resulting equation of motion reproduces the Aharonov--Casher interaction in the
non-relativistic limit.

The starting point is the component Lagrangian obtained after implementing the duality
identification between the symmetric combination of the chiral superfield and the
Kalb--Ramond superfield strength discussed in the previous section. The bosonic part of
the on-shell Lagrangian can be written as
\begin{equation}
\mathcal{L}=\mathcal{L}_1+\mathcal{L}_2+\mathcal{L}_3 ,
\end{equation}
with
\begin{align}
\mathcal{L}_1 &= \frac{\alpha_1}{4} M H_{\mu\nu}H^{\mu\nu}, \\
\mathcal{L}_2 &= -\frac{\alpha_1}{4}\epsilon^{\mu\alpha\beta\nu}
\tilde{G}_\mu H_{\alpha\beta}H_\nu ,\\
\mathcal{L}_3 &= \partial_\mu M \partial^\mu M
+ \frac16 G_{\mu\nu\lambda}G^{\mu\nu\lambda}
+ m \tilde{G}_\mu H^\mu .
\end{align}

To implement Lorentz symmetry breaking we adopt the ansatz
\begin{equation}
\rho = M = \text{const},
\end{equation}
for the real scalar component of the chiral superfield. This configuration freezes the
bosonic background while leaving the fermionic sector dynamical. Physically this
corresponds to a fixed Kalb--Ramond background selecting a preferred tensor structure
in spacetime.

Under this assumption the relevant dynamical contribution becomes the fermionic
Lagrangian.
Here $\Psi$ denotes the four-component Dirac spinor that is the Dirac lift of
the Kalb--Ramond spinor $\xi_\alpha$, identified via the duality
map~\eqref{eq:components} with the chiral spinor $\psi_\alpha$ of $S$; it is
distinct from the chiral spinor $\psi_\alpha$, which is set to zero by the
ansatz in the bosonic background.
The fermionic Lagrangian reads
\begin{equation}
\mathcal{L}_F =
i\alpha_1\bar{\Psi}\gamma^\mu\partial_\mu\Psi
+ \frac{\alpha_2}{2}\bar{\Psi}\sigma^{\mu\nu}F_{\mu\nu}\Psi .
\end{equation}

The Fayet--Iliopoulos sector introduces the auxiliary contribution
\begin{equation}
\mathcal{L}_{\text{aux}} =
4MD^2
-2\sqrt{2}\,\bar{\Psi}\Psi D .
\end{equation}

The equation of motion for the auxiliary field $D$ follows immediately:
\begin{equation}
\frac{\partial \mathcal{L}}{\partial D}=0
\quad \Rightarrow \quad
D=\frac{\sqrt{2}}{4M}\bar{\Psi}\Psi .
\end{equation}

Substituting this result back into the Lagrangian generates an effective fermionic
self-interaction and leads to the effective on-shell theory
\begin{equation}
\mathcal{L}_{\text{eff}} =
i\alpha_1\bar{\Psi}\gamma^\mu\partial_\mu\Psi
+ \frac{\alpha_2}{2}\bar{\Psi}\sigma^{\mu\nu}F_{\mu\nu}\Psi
- M \bar{\Psi}\Psi .
\end{equation}

The corresponding equation of motion for the fermionic field is therefore
\begin{equation}
\left(
i\gamma^\mu\partial_\mu
+ \frac{\alpha_2}{2\alpha_1}\sigma^{\mu\nu}F_{\mu\nu}
- M
\right)\Psi=0 .
\end{equation}

This equation has the structure of the Dirac equation for a neutral particle possessing
a magnetic dipole moment interacting with the electromagnetic field. Comparing with the
standard form
\begin{equation}
\left(
i\gamma^\mu\partial_\mu
+ \frac{\mu_{\text{eff}}}{2}\sigma^{\mu\nu}F_{\mu\nu}
- M
\right)\Psi=0 ,
\end{equation}
we identify the effective magnetic dipole moment generated dynamically by the model:
\begin{equation}
\mu_{\text{eff}}=\frac{\alpha_2}{2\alpha_1 M}.
\end{equation}

To make contact with the Aharonov--Casher interaction we now consider the
non-relativistic limit of the theory. Performing the standard Pauli reduction of the
Dirac equation, the Hamiltonian governing the dynamics of the two-component spinor
takes the form \cite{sakurai}
\begin{equation}
H =
\frac{1}{2M}
\left(
\mathbf{p}-\boldsymbol{\mu}_{\text{eff}}\times\mathbf{E}
\right)^2
-\boldsymbol{\mu}_{\text{eff}}\cdot\mathbf{B}.
\end{equation}

For a neutral particle in the presence of an external electric field the dominant
interaction term becomes
\begin{equation}
H_{AC} =
-\boldsymbol{\mu}_{\text{eff}}\cdot
(\boldsymbol{\sigma}\times\mathbf{E}),
\end{equation}
which is precisely the Hamiltonian responsible for the Aharonov--Casher phase \cite{Aharonov:1984zz,Cimmino:1989}.

Therefore the integration of the auxiliary $D$ field through the Fayet--Iliopoulos term
generates dynamically an effective dipole interaction. In the presence of the
Lorentz-violating Kalb--Ramond background this interaction produces the Aharonov--Casher
coupling in the non-relativistic regime, establishing the mechanism proposed in this
work.

% -------------------------------------------------------

% -----------------------------------------------SECAO V--------
%\section{Physical Implications and Phenomenological Bounds}
%\label{sec:discussion}
% -------------------------------------------------------
\section{Physical implications and phenomenological considerations}
\label{sec:discussion}

In this section we discuss the conceptual implications of the mechanism derived in the
previous section, its relation to earlier supersymmetric results on anomalous magnetic
dipole moments, and its connection with the Standard Model Extension (SME) framework
for Lorentz violation.

\subsection{Supersymmetry and the vanishing AMDM}

It is well known that in certain classes of supersymmetric gauge theories the anomalous
magnetic dipole moment (AMDM) of fermions vanishes when supersymmetry is exact \cite{FerraraRemiddi1974}.
This result was established in particular in the analysis of supersymmetric extensions of
Maxwell--Chern--Simons (MCS) theories, where cancellations between bosonic and
fermionic loop contributions enforce a zero AMDM.

At first sight this observation appears to forbid the emergence of the Aharonov--Casher
interaction in supersymmetric models, since the latter requires a nonvanishing dipole
coupling of the form
\begin{equation}
\bar{\psi}\sigma^{\mu\nu}F_{\mu\nu}\psi .
\end{equation}

The result obtained in the present work shows that this conclusion is not general.
In our model the dipole interaction does not arise from radiative corrections within a
Lorentz-invariant supersymmetric gauge sector. Instead, it emerges dynamically from the
integration of the auxiliary $D$ field associated with the Fayet--Iliopoulos term in the
presence of a Lorentz-violating background generated by the Kalb--Ramond field.

Therefore the vanishing of the AMDM in previous supersymmetric constructions must be
understood as model-dependent rather than a general theorem of exact supersymmetry.
Our result provides an explicit counterexample in which background supersymmetry is preserved
while a dipole interaction is nevertheless generated.

\subsection{Lorentz violation and supersymmetry}

The coexistence of Lorentz violation with exact supersymmetry raises an interesting
conceptual question. Since supersymmetry extends the Poincar\'e algebra, it might appear
that any breaking of Lorentz symmetry would necessarily induce a breaking of the
supersymmetry algebra as well.

However, it has been known since the work of Berger and Kosteleck\'y that supersymmetric
extensions of the Standard Model Extension (SME) can accommodate Lorentz-violating
backgrounds while preserving the supersymmetry algebra under certain conditions \cite{Kosta2002}.
In these constructions the Lorentz violation originates from vacuum expectation values
of tensor fields rather than explicit symmetry-breaking terms in the Lagrangian.

The mechanism realized in the present model falls naturally within this general picture.
The Kalb--Ramond field provides an antisymmetric tensor background
\begin{equation}
b_{\mu\nu}=\langle B_{\mu\nu}\rangle ,
\end{equation}
which breaks particle Lorentz invariance while preserving observer Lorentz invariance.
Because the identification between the chiral superfield sector and the Kalb--Ramond
superfield strength is implemented on-shell with $F=0$, the supersymmetry algebra
remains intact.

Similar constructions involving Lorentz-violating backgrounds in supersymmetric gauge
models have been explored in the literature by Helay\"el-Neto and collaborators, where
tensor backgrounds induce modified gauge interactions without necessarily destabilizing
the supersymmetric structure of the theory.

\subsection{Phenomenological bounds from the SME}

The Lorentz-violating background introduced in the present model can be interpreted
within the framework of the Standard Model Extension. In particular, the antisymmetric
tensor background generated by the Kalb--Ramond field contributes to the SME
coefficients in both the photon and fermion sectors.

In the fermionic sector the interaction generated in our model has the form
\begin{equation}
\mathcal{L}_{LV} =
\frac{\mu_{\text{eff}}}{2}
\bar{\psi}\sigma^{\mu\nu}F_{\mu\nu}\psi ,
\end{equation}
which corresponds to a tensor coupling associated with the SME coefficient
$H_{\mu\nu}$.

Current experimental constraints summarized in the SME data tables place strong bounds
on these coefficients. In the electron sector, laboratory experiments such as precision
spectroscopy and torsion-balance measurements constrain
\begin{equation}
|H_{\mu\nu}| \lesssim 10^{-27}\ \text{GeV}.
\end{equation}

Interpreting the effective dipole moment obtained in our model,
\begin{equation}
\mu_{\text{eff}}=\frac{\alpha_2}{2\alpha_1 M},
\end{equation}
as arising from the background tensor $b_{\mu\nu}$, these bounds translate into
corresponding constraints on the scale of the Kalb--Ramond background \cite{Kosta2011}. Such limits
are consistent with the expectation that Lorentz violation induced by tensor fields
originating in high-energy theories, such as string theory, should be suppressed by a
large mass scale \cite{Kosta1997,Kosta1998}.

\subsection{Possible extensions}

Several extensions of the mechanism presented here may be considered.

First, the construction could be generalized to non-Abelian gauge theories.
In such a scenario the dipole interaction would involve the non-Abelian field strength
tensor and could lead to new types of geometric phase effects for neutral fermions
coupled to non-Abelian gauge backgrounds.

A second natural direction concerns the dual topological phase known as the
He--McKellar--Wilkens effect, which describes the geometric phase acquired by an
electric dipole moving in a magnetic field. Since the mechanism derived in this work
generates dipole interactions dynamically from the supersymmetric sector, it would
be interesting to investigate whether a similar supersymmetric construction could
produce the dual electric dipole coupling required for the He--McKellar--Wilkens phase.

We leave these extensions for future investigation.

% ----------------------------------------------CONCLUSOES---------
\section{Conclusions}
\label{sec:conclusions}

In this work we have constructed a supersymmetric $N=1$, $D=4$ gauge model in which
the Aharonov-Casher interaction emerges dynamically in the presence of a
Lorentz-violating Kalb-Ramond background. The model combines a chiral superfield,
an abelian gauge superfield, and a Kalb-Ramond superfield coupled through a
Chern-Simons-type interaction and supplemented by a Fayet-Iliopoulos sector.

A key ingredient of the construction is the duality identification between the
symmetric combination $(S+S^\dagger)$ of the chiral superfield and the Kalb-Ramond
superfield strength $G$. This identification allows the antisymmetric tensor
background to be interpreted as a particular vacuum configuration in the chiral
sector, providing a natural mechanism for spontaneous Lorentz symmetry violation
while preserving the supersymmetry algebra.

Integrating out the auxiliary $D$ field associated with the Fayet-liopoulos term
generates dynamically an effective dipole interaction of the form
$\bar{\psi}\sigma^{\mu\nu}F_{\mu\nu}\psi$. The resulting equation of motion
coincides with the Dirac equation for a neutral fermion possessing a magnetic
dipole moment $\mu_{\text{eff}}=\frac{\alpha_2}{2\alpha_1 M}$, which leads, in the non-relativistic limit, to the Hamiltonian responsible for
the Aharonov-Casher phase.

Our result therefore provides an explicit realization of a supersymmetric theory
in which the dipole interaction required for the Aharonov-Casher effect arises
dynamically from the supersymmetric structure of the model rather than being
introduced phenomenologically. In particular, the mechanism relies on the
integration of the auxiliary $D$ field in the presence of a Lorentz-violating
Kalb-Ramond background.

The dipole operator arises from the supersymmetric FI sector rather
than being introduced phenomenologically.

From a conceptual perspective, the model also illustrates that the vanishing of
the anomalous magnetic dipole moment in certain supersymmetric gauge theories is
not a universal consequence of exact supersymmetry, but rather a feature of
specific Lorentz-invariant constructions. When Lorentz symmetry is spontaneously
broken by a tensor background, supersymmetry can remain intact while effective
dipole interactions are generated.

Finally, the framework presented here naturally connects with the Standard Model
Extension description of Lorentz violation, allowing the Kalb-Ramond background
to be interpreted in terms of SME coefficients and constrained by existing
experimental bounds.

Future research could fruitfully explore possible extensions to non-Abelian gauge theories, where similar mechanisms
could generate dipole interactions in non-Abelian backgrounds, and the
construction of supersymmetric models capable of producing the dual
He-McKellar-Wilkens phase associated with electric dipole couplings.

We hope that the mechanism discussed here may provide a useful framework for
exploring the interplay between supersymmetry, Lorentz violation, and geometric
quantum phases in field theory.

% -------------------------------------------------------

%\textit{[A ser desenvolvido. Esta seção resumirá os três resultados principais:
%construção do modelo, mecanismo da dualidade $S + S^\dagger \leftrightarrow
%\mathcal{G}$ como gerador de violação de Lorentz com SUSY intacta, e emergência
%do efeito AC via termo FI.]}

% -------------------------------------------------------
\section*{Acknowledgments}
C.A.S.A.\ is supported by grants No.~309553/2021-0 and No.~420854/2025-8
(CNPq) and by Project UNI-00210-00230.01.00/23 (FUNCAP).

% -------------------------------------------------------

\end{document}